\documentclass[aps,twocolumn,superscriptaddress]{revtex4}

\usepackage{graphicx}
\usepackage{amsmath}
\usepackage{amssymb}

\begin{document}
\title{Envelope-driven recollisions triggered by an elliptically polarized pulse}

\author{J. Dubois}
\affiliation{Aix Marseille Univ, CNRS, Centrale Marseille, I2M, Marseille, France}
\author{C. Chandre}
\affiliation{Aix Marseille Univ, CNRS, Centrale Marseille, I2M, Marseille, France}
\author{T. Uzer}
\affiliation{School of Physics, Georgia Institute of Technology, Atlanta, Georgia 30332-0430, USA}

\begin{abstract}
Increasing ellipticity usually suppresses the recollision probability drastically. 
In contrast, we report on a recollision channel with large return energy and a substantial probability, regardless of the ellipticity. The laser envelope plays a dominant role in the energy gained by the electron, and in the conditions under which the electron comes back to the core. We show that this recollision channel efficiently triggers multiple ionization with an elliptically polarized pulse.
\end{abstract}

\maketitle

Recollision~\citep{Corkum1993, Schafer1993, Corkum2011} is the fundamental building block of many imaging techniques for the real-time motion of electrons because recolliding electrons probe the structural dynamics~\citep{Gaffney2007, Smirnova2009, AbuSahma2018} of the target atom or molecule, giving rise to a variety of highly nonlinear and nonperturbative phenomena such as high harmonic generation (HHG), above-threshold ionization (ATI) and nonsequential multiple ionization (NSMI). 
The conventional semiclassical scenario of the recollisions~\citep{Corkum1993,Schafer1993} is split into three distinct steps where: (i) The electron tunnel-ionizes through the potential barrier induced by the laser field, (ii) travels in the laser field and gains energy, then returns to the core, and, in the case of nonsequential double ionization (NSDI), (iii)  exchanges its excess energy with an electron of the parent ion and both ionize. This recollision picture can be easily understood in absence of a pulse envelope in the dipole approximation: For an elliptically polarized (EP) laser field $\mathbf{E}(t) = E_0 [\hat{\mathbf{x}} \cos (\omega t) + \xi \hat{\mathbf{y}} \sin (\omega t )]$, given an ionization time $t_{\rm i}$, the position of the electron after ionization in the strong field approximation (SFA) is given by (atomic units are used unless otherwise stated)
\begin{equation}
\label{eq:momentum_recollision_condition}
\mathbf{r} ( t ) = \mathbf{r} (t_{\rm i}) + \left[ \mathbf{p}(t_{\rm i}) - \mathbf{A}(t_{\rm i}) \right] (t-t_0) + [\mathbf{E}(t) - \mathbf{E}(t_{\rm i})]/\omega^2 .
\end{equation}
The amplitude of the laser field is $E_0$, its ellipticity is $\xi$ and its frequency is $\omega$. The position and the kinetic momentum of the electron are $\mathbf{r}$ and $\mathbf{p}$, respectively. The vector potential is $\mathbf{A}(t)$ such that $\mathbf{E}(t) = - \partial \mathbf{A}(t)/\partial t$. At ionization, the electron is on the potential barrier, and $\mathbf{p}(t_{\rm i}) = \boldsymbol{0}$. For $\xi = 0$, the electron leaves the ionic core with no sideways momentum drifts, and returns close to the core at time $t_{\rm r}$, i.e., $\mathbf{r}(t_{\rm r}) \approx \mathbf{r}(t_{\rm i})$, due to the laser oscillations $\mathbf{E}(t)/\omega^2$. For $\xi > 0$, if $\mathbf{p}(t_{\rm i}) = \boldsymbol{0}$, strictly speaking the electron never comes back to the core, i.e., there are no times $t_{\rm i}$ and $t_{\rm r}$ such that $\mathbf{r}(t_{\rm r}) = \mathbf{r}(t_{\rm i})$ in Eq.~\eqref{eq:momentum_recollision_condition} because of the non-vanishing drift momentum $- \mathbf{A}(t_{\rm i})$ which pushes the electron away from the core~\citep{Corkum1993}. In this framework, the electron can only recollide and trigger NSDI when the laser field is linearly polarized (LP).

\par
In the conventional three-step scenario, the recollision picture can, however, be extended to near LP-fields by taking into account that after ionization, the electron is near the potential barrier. The initial momentum along the transverse direction to the laser field $p_{\perp}$ is distributed~\citep{Ammosov1986,Arissian2010} as $\propto \exp (- p_{\perp}^2 \sqrt{2 I_p} / E_0)$, with $I_p$ denoting the ionization potential of the atom. The initial momentum of the electron can compensate its initial drift after ionization, and recollisions become possible~\citep{Wang2010_PRL,Budil1993,Moller2012}. At the threshold ellipticity~\citep{Moller2012} $\xi_{\rm th} \approx \omega I_p^{-1/4} / \sqrt{E_0}$ (which is $\xi_{\rm th} \approx 0.5$ for the parameters we use here) and beyond, the initial momentum $p_{\perp} \approx \xi E_0/\omega$ necessary to compensate the drift momentum is poorly weighted and the probability that the electron returns to the core drops off drastically. As a consequence, for ellipticities $\xi > \xi_{\rm th}$, the NSDI probability drops off compared to its value for $\xi = 0$~\citep{Wang2010_PRL}. Moreover, if the electron returns to the core, the energy it gains during the recollision is
\begin{equation}
\label{eq:energy_gain}
\Delta \mathcal{E} = \kappa \mathrm{U}_p (1 - \xi^2) ,
\end{equation}
with $\mathrm{U}_p = E_0^2/4\omega^2$ denoting the ponderomotive energy and $0 \leq \kappa \lesssim 3.17$. Hence, the recolliding electrons, if any, do not bring back enough energy from the laser field at ellipticities $\xi$ close to $1$ to trigger NSDI. Consequently, the three-step model in absence of laser envelope predicts that the recollisions and the enhancement of DI probability are suppressed in elliptically polarized light. 
 
\begin{figure*}
\includegraphics[width=\textwidth]{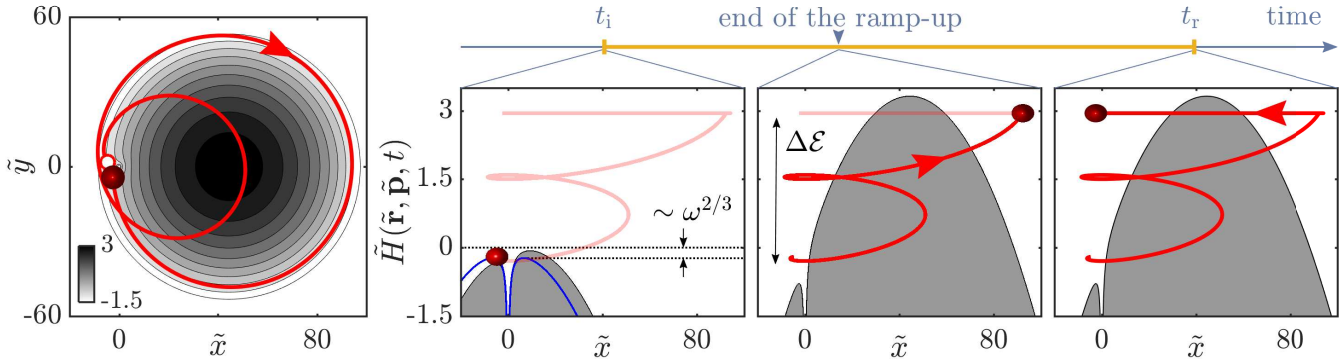}
\caption{Typical recollision with high return energy in a CP field. The laser envelope is trapezoidal 2--4--2 (see text), the laser intensity is $I = 8 \times 10^{14} \; \mathrm{W}\cdot \mathrm{cm}^{-2}$. The electron is initialized with energy $- I_p = - 5.8 \; \mathrm{eV}$. Left panel: Trajectory in the rotating frame $(\tilde{x},\tilde{y})$. The white and red circles are the position of the electron at the ionization and return time, respectively. The color shows the value of the zero-velocity surface during the plateau. Right panels: The trajectory projected along the position and energy of Hamiltonian~\eqref{eq:Hamiltonian_RF} at ionization $t_{\rm i} \approx 0.3 T$, the end of the ramp-up $2T$ and return $t_r \approx 2.6 T$ (from left to right). The gray surface is the classically forbidden region for $\tilde{y} = 0$ [i.e., $H (\tilde{\mathbf{r}},\tilde{\mathbf{p}} , t) < \mathcal{Z} (\tilde{\mathbf{r}} , t)$]. The blue curve is the zero-velocity surface at time $t=0$ for $\tilde{y} =0$. The saddle point is the local maximum of the zero-velocity surface for negative $\tilde{x}$. Distances and energies are in $\mathrm{a.u.}$}
\label{fig:scenario_illustration}
\end{figure*} 
 
\par
Remarkably, we find that the presence of an envelope drastically affects these conclusions: In this Letter, we unveil a highly probable recollision channel with large return energy by accounting for the effects of the pulse envelope $f(t)$. The mechanism for this recollision channel is depicted in Fig.~\ref{fig:scenario_illustration}. The competition between the Coulomb force and the laser field induces ionization early after the laser field is turned on. As the electron leaves the core region, the amplitude of the vector potential is small, and therefore its sideways momentum drift $- \mathbf{A}(t_{\rm i}) \propto \hat{\mathbf{y}} \xi f(t_{\rm i})$ can be compensated by its momentum $\mathbf{p}(t_{\rm i})$ [see Eq.~\eqref{eq:momentum_recollision_condition}], and this, regardless of the ellipticity of the laser. This recollision channel becomes highly probable if the ionization is over the barrier. The conditions under which this recollision channel is highly probable lead to the enhancement of the DI probability for specific atoms only, driven by circularly polarized (CP) pulses~\citep{Fittinghoff1994, Guo1998, Gillen2001, Mauger2010_PRL} and nearly-circularly polarized pulses. Furthermore, we show that this recollision channel and the return of the electron can be understood using the SFA, contrary to the prediction of the standard three-step model as summarized in Ref.~\citep{Corkum2011}.

\par
We demonstrate the existence of this recollision channel, which will be referred to as envelope-driven recollisions, for the least favorable case in the recollision scenario in absence of pulse envelope, namely the circularly polarized (CP) case ($\xi = 1$). We describe the electron dynamics in a rotating frame (RF). In the RF, the CP field is unidirectional. The position and the momentum of the electron are $\tilde{\mathbf{r}} = \mathbf{R}(t) \mathbf{r}$ and $\tilde{\mathbf{p}} = \mathbf{R}(t) \mathbf{p}$, with $\mathbf{R}(t)$ the rotation matrix of angle $\omega t$ in the polarization plane. The classical single-active electron Hamiltonian, in the dipole approximation and in the three-dimensional space with basis vectors $(\hat{\mathbf{x}},\hat{\mathbf{y}},\hat{\mathbf{z}})$, reads~\cite{Peng2015}
\begin{eqnarray}
\label{eq:Hamiltonian_RF}
H ( \tilde{\mathbf{r}}, \tilde{\mathbf{p}}, t) &=& \dfrac{|\tilde{\mathbf{p}}|^2}{2} - \omega \tilde{\mathbf{p}} \cdot  \hat{\mathbf{z}} \times \tilde{\mathbf{r}} + V (\tilde{\mathbf{r}}) + \tilde{x} E_0 f(t) ,
\end{eqnarray}
where the term $- \omega \tilde{\mathbf{p}} \cdot  \hat{\mathbf{z}} \times \tilde{\mathbf{r}}$ is the Coriolis potential. We use a laser wavelength of $780\;\mathrm{nm}$ (corresponding to $\omega = 0.0584 \; \mathrm{a.u.}$), a laser intensity of $8 \times 10^{14} \; \mathrm{W}\cdot\mathrm{cm}^{-2}$ (corresponding to $E_0= 0.151 \; \mathrm{a.u.}$), and the soft Coulomb potential~\citep{Javanainen1988} $V (\mathbf{r}) = - (|\mathbf{r}|^2 + a^2)^{-1/2}$ with $a = 0.26$. At each time, the Hamiltonian is $H (\tilde{\mathbf{r}},\tilde{\mathbf{p}},t) \geq \mathcal{Z} (\tilde{\mathbf{r}},t)$, with $\mathcal{Z} (\tilde{\mathbf{r}}, t) = -\omega^2 |\tilde{\mathbf{r}} \times \hat{\mathbf{z}} |^2/2 + V(\tilde{\mathbf{r}}) + \tilde{x} E_0 f(t)$. In the right panels of Fig.~\ref{fig:scenario_illustration}, the gray surfaces are classical forbidden regions corresponding to the condition $H (\tilde{\mathbf{r}},\tilde{\mathbf{p}},t) < \mathcal{Z} (\tilde{\mathbf{r}},t)$. The boundaries of the gray surfaces are the zero-velocity surface~\citep{Hill1878} corresponding to the condition $H (\tilde{\mathbf{r}} , \tilde{\mathbf{p}} , t) = \mathcal{Z} (\tilde{\mathbf{r}} , t)$. On the zero-velocity surface, $\tilde{\mathbf{p}} = \omega \hat{\mathbf{z}} \times \tilde{\mathbf{r}}$. In the adiabatic approximation, there exists three fixed points of the dynamics: at the top of the zero-velocity surface and mainly due to the laser interaction [$\tilde{\mathbf{r}} \approx \hat{\mathbf{x}} f(t) E_0/\omega^2$], around the origin and mainly due to the soft Coulomb potential ($\tilde{\mathbf{r}} \approx \boldsymbol{0}$), and at the saddle point $\tilde{\mathbf{r}}^{\star} = \hat{\mathbf{x}} \tilde{x}^{\star}$ with $\tilde{x}^{\star}$ solution of the equation $\omega^2 \tilde{x} - \partial V (\tilde{x}^{\star} \hat{\mathbf{x}})/\partial \tilde{x} = f(t) E_0$ which is due to the competition between the Coulomb potential, the laser interaction and the Coriolis potential. At time $t$, the energy of the saddle point is denoted $\mathcal{Z}^{\star} (t) = \mathcal{Z} ( \tilde{\mathbf{r}}^{\star},t)$.

\par
First, we consider a constant laser envelope with $f=1$. Hamiltonian~\eqref{eq:Hamiltonian_RF} is conserved in time, and its value is the Jacobi constant $\mathcal{K} = H ( \tilde{\mathbf{r}}, \tilde{\mathbf{p}})$. We apply the recollision picture in which the electron ionizes at time $t_{\rm i}$ and returns at time $t_{\rm r}$ such that $\tilde{\mathbf{r}}(t_{\rm i}) \approx \tilde{\mathbf{r}}(t_{\rm r}) \approx \boldsymbol{0}$ close to the core. Between time $t_{\rm i}$ and $t_{\rm r}$, the electron is in the continuum. Because of the Jacobi constant, $H ( \tilde{\mathbf{r}} (t_{\rm i}), \tilde{\mathbf{p}}(t_{\rm i})) = H (\tilde{\mathbf{r}} (t_{\rm r}), \tilde{\mathbf{p}}(t_{\rm r}))$, and consequently the return energy of the electron is $| \tilde{\mathbf{p}}(t_{\rm r})|^2/2 \approx | \tilde{\mathbf{p}}(t_{\rm i})|^2/2$. Hence, the electron does not gain energy during its excursion in the continuum when it is driven by CP light even in the presence of the Coulomb potential, in agreement with Eq.~\eqref{eq:energy_gain}.

\par
When the laser envelope is taken into account, however, Hamiltonian~\eqref{eq:Hamiltonian_RF} is no longer conserved and the energy of the recolliding electron can vary during its excursion. We consider $f(t)$ to be a 2--4--2 trapezoidal envelope (2 laser cycles ramp-up, 4 laser cycles plateau and 2 laser cycles ramp-down), unless stated otherwise. In Fig.~\ref{fig:scenario_illustration}, we show a typical recollision in CP fields seen in the rotating frame of an electron initialized with an energy $H (\tilde{\mathbf{r}} , \tilde{\mathbf{p}} , 0) = - I_p$. When the laser is turned off ($f = 0$), the motion of the electron is bounded. The zero-velocity surface is the blue curve in Fig.~\ref{fig:scenario_illustration}. The maximum of the zero-velocity surface (top of the blue curve) is $\mathcal{Z}^{\star}(0) \approx - (3/2) \omega^{2/3}$, i.e., the Coulomb barrier is decreased in the RF since this framework naturally includes nonadiabatic effects~\citep{Mauger2014_JPB2}. In a zeroth-order approximation, we consider two situations: If $I_p < (3/2) \omega^{2/3}$, as it is the case for the ionization potential used here, the electron is initially above the classical forbidden region, and a low laser intensity is sufficient to tear the electron off the core almost instantaneously, so $t_{\rm i} \approx 0$. If $I_p > (3/2) \omega^{2/3}$, the electron is initially topologically bounded by the zero-velocity surface. The ionization is not instantaneous, so $t_{\rm i} > 0$. For $t < t_{\rm i}$, the time-dependent term $\tilde{x} E_0 f(t)$ in Hamiltonian~\eqref{eq:Hamiltonian_RF} is very small, and therefore the energy is almost conserved $H (\tilde{\mathbf{r}},\tilde{\mathbf{p}},t) \approx - I_p$. For increasing time, the envelope increases and the energy of the saddle point decreases as $\mathcal{Z}^{\star}(t) \approx \mathcal{Z}^{\star}(0) - f(t) E_0 \omega^{- 2/3}$ for $f(t) \ll 1$. The ionization channel is opened when the energy of the saddle point is the same as the energy of the electron $\mathcal{Z}^{\star}(t_{\rm i}) = - I_p$. Quantum mechanically, earlier ionization times can be accessible through tunneling. In any case, the electron ionizes close to the zero-velocity surface, and the electron initial kinetic energy $\mathcal{E}_{\rm i}$ is small compared to the ponderomotive energy $\mathrm{U}_p$.

\par
After ionization, at time $t > t_{\rm i}$, the electron is outside the core region and the time-dependent term $\tilde{x} E_0 f(t)$ in Hamiltonian~\eqref{eq:Hamiltonian_RF} starts varying significantly compared to $\mathrm{U}_p$. If $t_{\rm i}$ is relatively small [$f(t_{\rm i}) \ll 1$], the effective transverse potential vector $f(t_{\rm i}) E_0/\omega \hat{\mathbf{y}}$ is also small, and it can be compensated by the momentum of the electron at ionization. This prevents the electron from drifting away from the core, which is the case if the ionization occurs when $f ( t_{\rm i} ) \sim 1$. During its excursion outside the core region, the electron spins around the zero-velocity surface. As we show below, the energy of the recolliding electron increases by a quantity of order $\mathrm{U}_p$ during the ramp-up. Therefore, the electron can populate high energy states. In particular, regions of phase space where invariant structures can bring back this electron to the core region~\citep{Mauger2010_PRL, Wang2010_PRL, Kamor2013}. During the plateau, the electron dynamics evolves on a constant energy surface given by the Jacobi constant. 

\par
The ionization time $t_{\rm i}$ determines whether or not the electron comes back: If $t_{\rm i}$ is too large, there is a low probability that the momentum of the electron at ionization can compensate the sideways momentum drift $- \mathbf{A}(t_{\rm i})$. Two parameters influence considerably the ionization time $t_{\rm i}$: the electron initial energy $- I_p$ and the laser frequency $\omega$. We find that the electron comes back to the core after ionizing over the barrier only if $I_p < I_c$, where $I_c \approx 1.85 \omega^{2/3}$. This expression is derived using a reduced model~\citep{Dubois2018_PRE} (see Supplemental Material~\citep{SuppMat} for details). For $\omega = 0.0584 \; \mathrm{a.u.}$, the critical value is $I_c \approx 7.4 \; \mathrm{eV}$. If $I_p > I_c$, the vector potential of the electron at ionization is too large and cannot be compensated by the electron, and it drifts away from the core without recolliding. We notice that $I_c/\mathcal{Z}^{\star} (0) \approx - 1.2$, which shows that the critical ionization potential is right below the saddle point energy before the laser field is turned on. Conversely, for a given $I_p$, the laser frequency can be tuned for allowing recollisions in CP fields, as observed numerically in Ref.~\citep{Chen2017}. Quantum mechanically, the wave packet can ionize before their classical counterparts, and therefore increasing the range of $I_p$ which allows recollisions. 

\par
In the RF with $\xi \neq 1$, the recollision picture is not so clear since the saddle moves in time even with a fixed laser envelope. The scenario, however, works the same way: The electron initiated with a sufficiently large energy can ionize over the barrier early after the laser field is turned on. After ionization, the initial momentum of the electron is relatively small and compensates the sideways momentum drift in Eq.~\eqref{eq:momentum_recollision_condition}, such that $\mathbf{p}(t_{\rm i}) - \mathbf{A}(t_{\rm i}) \approx \boldsymbol{0}$. The electron does not quickly drift away from the core. Then, the electron travels in the continuum. During its excursion in the continuum, the energy gained by the electron for a slowly varying envelope is
\begin{equation}
\label{eq:Difference_energy_envelope}
\Delta \mathcal{E} \approx 2 \mathrm{U}_p \left[ f(t_{\rm r})^2 - f(t_{\rm i})^2 \right] + 2 \mathrm{U}_p \left( 1 - \xi^2 \right)  g (t_{\rm i},t_{\rm r}) ,
\end{equation}
where $g(t_{\rm i},t_{\rm r})$ is explicitly derived from the SFA; it depends on the parameters of the envelope and on the frequency of the laser, but is independent of $\xi$ (see Supplemental Material~\citep{SuppMat} for its analytic expression). Therefore, at high ellipticities, the recolliding electron gains energy mostly through the laser envelope [first term of the right hand-side of Eq.~\eqref{eq:Difference_energy_envelope}]. We observe that the energy of the electron can increase by an order of $\mathrm{U}_p$ with the variations of the pulse envelope. In contrast, at low ellipticities, the recolliding electron gains energy mostly through the sub-cycle oscillations of the laser [second term of the right hand-side of Eq.~\eqref{eq:Difference_energy_envelope}]. 

\begin{figure}
\includegraphics[width=0.5\textwidth]{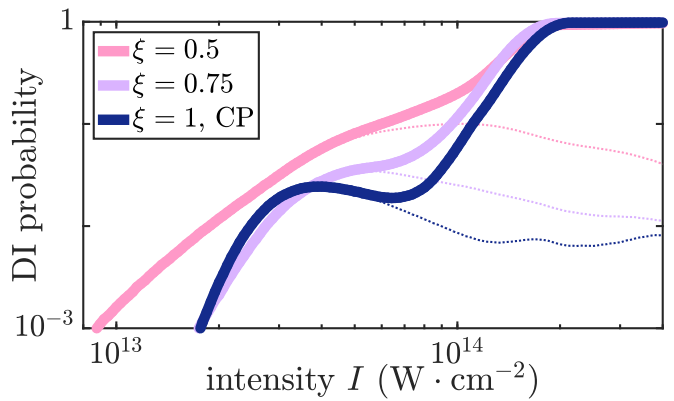}
\caption{Double ionization probability curves as a function of the laser intensity. The thin dotted curves are the NSDI probability curves.}
\label{fig:DI_probability}
\end{figure}

\par
We use this classical recollision picture to analyze NSDI with an elliptically polarized pulse~\citep{Wang2010_PRL,Mauger2010_PRL}. Figure~\ref{fig:DI_probability} shows the double ionization probability curves of $\mathrm{Mg}$ with Hamiltonian~\citep{Mauger2010_PRL}
\begin{equation}
\label{eq:Hamiltonian_NSDI}
\mathcal{H}  = \sum_{k=1}^2 \left[ \dfrac{|\mathbf{p}_k|^2}{2} + V (\mathbf{r}_k) + \mathbf{r}_k \cdot \mathbf{E}(t) \right] + \dfrac{1}{\sqrt{|\mathbf{r}_1 - \mathbf{r}_2|^2 + 1}} ,
\end{equation}
as a function of the laser intensity. The electric field in the LF is $\mathbf{E}(t) = E_0 f(t) [\hat{\mathbf{x}} \cos (\omega t ) + \hat{\mathbf{y}} \xi \sin(\omega t )]$. The electrons are initiated in the ground state of energy $\mathcal{E}_g = - 0.83 \; \mathrm{a.u.}$~\citep{Mauger2010_PRL} using a microcanonical distribution of the initial conditions. In Fig.~\ref{fig:DI_probability}, the NSDI probability is also represented. A NSDI is a DI triggered by recollisions. A DI is counted as NSDI if there exist $t_{\rm i}$ and $t_{\rm r}$ such that the recolliding electron (labeled $k=1$) leaves the core at time $t_{\rm i}$, i.e., $|\mathbf{r}_1 (t_{\rm i})| > R$, goes outside the core region, then returns to the core while the other electron remained bounded, i.e., $|\mathbf{r}_1 (t_{\rm r})| < R$ and $|\mathbf{r}_2 (t_{\rm r})| < R$, and both ionize. We used $R = 5 \; \mathrm{a.u.}$ in our simulations.

\begin{figure}
\centering
\includegraphics[width=.5\textwidth]{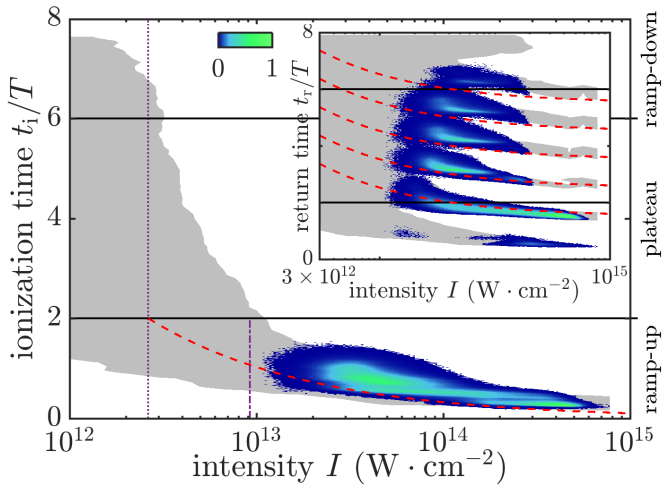}
\caption{Distributions of the ionization time $t_{\rm i}$ and the return $t_{\rm r}$ of the recollisions of Hamiltonian~\eqref{eq:Hamiltonian_NSDI} leading to NSDI for $\mathrm{Mg}$ and CP pulses ($\xi = 1$). The gray areas are where electrons undergo a recollision but do not lead to NSDI. The red dashed curve is our prediction of the ionization time of the outermost electron $t_{\rm i}$~\citep{SuppMat} for $I_p = 0.28 \; \mathrm{a.u.}$ In the inset, the red dotted curves are $t_{\rm r} = t_{\rm i} + (n + 1/2) T$ with $n \in \mathbb{N}^{\star}$. The vertical dotted curve is where we expect over-the-barrier ionization, and the vertical dash-dotted curve is where we expect envelope-driven recollisions to gain enough energy to trigger NSDI~\citep{SuppMat}.}
\label{fig:ionization_return_time}
\end{figure}

\par
In Fig.~\ref{fig:DI_probability}, the DI probability curves for $\mathrm{Mg}$ clearly exhibit knee structures as a function of the laser intensity for $\xi = 0.5$, $0.75$, $1$, signature of a recollision mechanism~\citep{Mauger2010_PRL}. In order to determine the type of recollisions responsible for this knee structure, which persists at high ellipticities, we examine the ionization and return times of the recollisions leading to NSDI. Figure~\ref{fig:ionization_return_time} shows the ionization and return times of all recollisions (gray areas), and the distributions of ionization and return times of the recollisions leading to NSDI (color scale) for $\mathrm{Mg}$. We observe that all recolliding electrons contributing to NSDI ionize during the ramp-up of the laser pulse. Recolliding electrons which ionize during the plateau or at the end of the ramp-up cannot bring back sufficiently energy to the core to trigger NSDI [see Eq.~\eqref{eq:energy_gain}]. As a consequence, an absence of NSDI is observed for intensities smaller than $I_{\min} \approx 9 \times 10^{12} \; \mathrm{W}\cdot\mathrm{cm}^{-2}$ (vertical dash-dotted curve) for high ellipticities, despite the presence of recolliding electrons. We observe layers for the return times of the recolliding electrons around $t_{\rm r} \approx t_{\rm i} + (n + 1/2)T$ with $n \in \mathbb{N}^{\ast}$. The probability of return is roughly equally distributed between $n = 1,2,3,4$, in contrast to the predictions of Ref.~\citep{Fu2012}. At recollision time $t_{\rm r}$, the energy of the recolliding electron is $\mathcal{E}_{\rm r} = \mathcal{E}_{\rm i} + \Delta \mathcal{E} \approx \Delta \mathcal{E}$. After the electron exchanges its energy with an electron which remained bounded, and if the energy the electron gained during the recollision $\Delta \mathcal{E}$ is sufficiently large, both ionize. 

\par
\begin{figure}
	\centering
	\includegraphics[width=.5\textwidth]{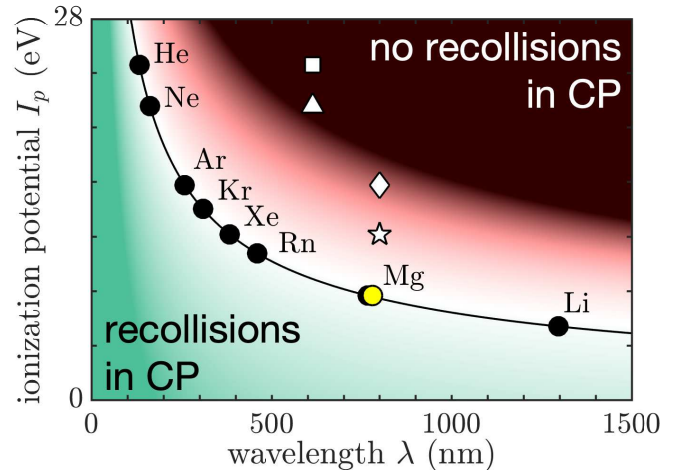}
	\caption{Phase diagram of the existence of envelope-driven recollisions. The black line is the critical ionization potential $I_c \approx 1.85 \; \omega^{2/3}$. The white (resp. yellow) markers are where no knee (resp. a knee) is observed in the double ionization probability curves. The experimental measurements for $\mathrm{He}$ ($\square$) and $\mathrm{Ne}$ ($\bigtriangleup$) at $\lambda = 614 \; \mathrm{nm}$ are reported in Ref.~\cite{Fittinghoff1994}; for $\mathrm{Ar}$ ($\diamond$) and $\mathrm{Xe}$ ($\star$) at $\lambda = 800 \; \mathrm{nm}$ are reported in Ref.~\cite{Guo1998}; for $\mathrm{Mg}$ ($\circ$) at $\lambda = 780 \; \mathrm{nm}$ are reported in Ref.~\cite{Gillen2001}. The color code is the guiding-center energy of the electron at the saddle point $\mathcal{E}$ at time $t_{\rm i}$: red is where $\mathcal{E} = 2 \; \mathrm{a.u.}$, white is where $\mathcal{E} = 0$ (critical ionization potential) and green is where $\mathcal{E} = - 1 \; \mathrm{a.u.}$}
	\label{fig:phase_diagram}
\end{figure}

While the recollisions from the conventional scenario do not persist at high ellipticities, or at least, are not efficient due to the lack of energy boost [see Eq.~\eqref{eq:energy_gain}], envelope-driven recollisions manifest themselves in an efficient way regardless of the laser ellipticity and with a substantial probability, provided that $I_p$ is smaller than a critical value, e.g., $I_p \lesssim 1.85 \omega^{2/3}$ for CP. Under this condition, which corresponds to the green area in Fig.~\ref{fig:phase_diagram}, manifestations of the recollisions, such as for instance in NSDI, are probable. We observe that the experimental measurements for which a knee structure is observed in CP pulses (see Refs.~\citep{Fittinghoff1994, Guo1998, Gillen2001}) agree well with the conditions under which the probability of envelope-driven recollisions is substantial. Therefore, manifestations of the recollisions can also be observed at high ellipticities for specific target atoms and laser wavelengths, and therefore triggering NSDI, HHG and ATI. We observe that, by decreasing the laser wavelength, the probability of envelope-driven recollisions becomes substantial for larger ionization potentials, and therefore manifestations of the recollisions becomes measurable for other atoms than $\mathrm{Mg}$, such as it was noticed in Ref.~\citep{Chen2017}.

\par
In summary, we have demonstrated the conditions under which a large portion of electrons can both ionize and return to their parent ion regardless of the ellipticity of the laser pulse. To recollide at high ellipticity, the electron needs to ionize early after the laser field is turned on to benefit from the boost by the CP field and not to drift away from the core. We have shown that the probability the electrons undergo an envelope-driven recollision is particularly large when $I_p \lesssim 1.85 \omega^{2/3}$. During its excursion in the continuum, the electron gains energy dominantly from the envelope of the driving laser, and can exceed $2 \mathrm{U}_p$, allowing for a return energy $\mathcal{E}_{\rm r} = \mathcal{E}_{\rm i} + \Delta \mathcal{E}$ in excess of $2 \mathrm{U}_p$. 

\par
We thank Simon A. Berman and Fran{\c c}ois Mauger for helpful discussions. The project leading to this research has received funding from the European Union's Horizon 2020 research and innovation program under the Marie Sk\l{}odowska-Curie grant agreement No.~734557. T.U. acknowledges funding from the NSF (Grant No.~PHY1602823).

%


\end{document}